\documentclass{webofc}
\usepackage[varg]{txfonts}   
\usepackage[autostyle, english = american]{csquotes}
\usepackage{hyperref}
\MakeOuterQuote{"}

\begin{document}

\title{Coffea-casa: an analysis facility prototype}

\author{\firstname{Matous} \lastname{Adamec}\inst{1}
\and
\firstname{Garhan} \lastname{Attebury}\inst{1}
\and
\firstname{Kenneth} \lastname{Bloom}\inst{1}
\and
\firstname{Brian} \lastname{Bockelman}\inst{2}
\and
\firstname{Carl} \lastname{Lundstedt}\inst{1}
\and
\firstname{Oksana} \lastname{Shadura}\inst{1}
\and
\firstname{John} \lastname{Thiltges}\inst{1}
}

\institute{University of Nebraska-Lincoln, Lincoln, NE 68588 
\and
          Morgridge Institute for Research, 330 N. Orchard Street, Madison, WI 53715
          }

\abstract{
Data analysis in HEP has often relied on batch systems and event loops; users are given a non-interactive interface to computing resources and consider data event-by-event.  The "Coffea-casa" prototype analysis facility is an effort to provide users with alternate mechanisms to access computing resources and enable new programming paradigms.  Instead of the command-line interface and asynchronous batch access, a notebook-based web interface and interactive computing is provided.  Instead of writing event loops, the column-based Coffea library is used.

In this paper, we describe the architectural components of the facility, the services offered to end users, and how it integrates into a larger ecosystem for data access and authentication.
}
\maketitle
\section{Introduction}
\label{intro}
The ultimate purpose of computing for high-energy physics is to serve physicists in minimizing "time to insight".  Physicists want quick access to data and incremental access to results, so that outputs can be examined almost as soon as the computational job is submitted.  The tools for this
must be approachable for inexperienced users while providing value to power users on the largest data analyses. 

Traditionally, computing for physics data analysts has been delivered through "analysis facilities", which provide physicists with a login account that has access to disk storage hosting experiment datasets and a batch-processing system for user jobs that take the data as input.  The LPC Central Analysis Facility~\cite{bib:lpc} at Fermilab is an example that has served Compact Muon Solenoid (CMS) \cite{bib:cms} physicists well during Run~1 and Run~2 of the Large Hadron Collider (LHC).  It has over 11,000 batch slots available to users, who can access data over 10 PB of local storage.  These facilities are typically used by physicists for a final round of data reduction, converting events in a 60~kB/event "MiniAOD" format \cite{bib:miniaod} at 25~Hz to an analysis-specific 1~kB/event ntuple format that can later be processed at 1~kHz to make plots.

However, this model of an analysis facility -- and of analysis as a computational task -- will face challenges during the upcoming High-Luminosity LHC (HL-LHC) era.  With 5 times increased trigger rates compared to recent experience, the number of events needed to be processed for an analysis will increase by about a factor of thirty.  To keep the cost of storage under control, most analyses will have to switch from MiniAOD to a much smaller data format, a "NanoAOD"~\cite{bib:nanoaod} of only 2~kB/event, but this format will likely be missing data elements necessary for any given analysis.  New approaches to analysis facilities are thus increasingly urgent.

At the same time, new tools for data analysis have become available.  The Python ecosystem provides a diverse suite of analysis tools that are in broad use in the scientific community~\cite{bib:sci_comp_py}.  Many scientists are taking advantage of Jupyter notebooks~\cite{bib:jupyter_note}, which provide a straightforward way of documenting a data analysis.  Column-wise data analysis~\cite{bib:awkward}, in which a single operation on a vector of events replaces calculations on individual events serially, is  seen as a way to for the field to take advantage of vector processing units in modern CPUs, leading to significant speed-ups in throughput.  Also, declarative programming paradigms~\cite{bib:rdataframe} can make it simpler for physicists to intuitively code their analysis.

The challenges ahead and the new tools available have led us to develop a prototype  analysis facility which we have named "Coffea-casa" \cite{bib:coffeacasa}.  The name is inspired by the Coffea framework~\cite{bib:coffea} which is implemented using the Python language. Coffea and Python, with its package ecosystem and commodity big data technologies, improve time-to-insight, scalability, portability, and reproducibility of analysis.  The Coffea-casa facility is built to natively support that framework.  

In Coffea-casa, the user is presented with a Jupyter notebook interface that can be populated with code from a Git repository.  When the notebook is executed, the processing automatically scales out to resources on the Nebraska CMS Tier-2 facility, giving the user transparent interactive access to a large computing resource.  The facility has access to the entire CMS data set, thanks to the global data federation and local caches.  An important feature is access to a ``column service”; if a user is working with a compact data format (such as a  NanoAOD) that is missing a data element that the user needs, the facility can be used to serve that “column” from a remote site.  This allows only the compact data formats to be stored locally and augmented only as needed, a critical strategy for CMS to control the costs of storage in the HL-LHC era.  Ultimately, Coffea-casa provides interactivity, low-latency data access, reusability, and easy deployment.

In this paper we describe the design of Coffea-casa and its deployment at Nebraska, early measurements of performance, and future directions.


\section{Analysis Facility Motivation and Design}

\subsection{Analysis Facilities In Context}
\label{sec-1}
The Nebraska Tier-2 facility~\cite{bib:t2comp} has over 13,000 compute cores managed by HTCondor~\cite{bib:htcondor} along with 13~PB of data storage managed by HDFS; the facility is primarily for use by the CMS experiment.  The facility's resources are managed via a batch system that is only accessible through a grid-enabled gateway. The normal usage for the site is asynchronous; users could potentially wait hours for their first results.  We desire to provide synchronous interfaces, such as web-based Jupyter notebooks, that have "instant" results.  Unfortunately, "instant" is impossible given finite compute resources.  Instead, we aim to provide instant \textit{first} results through modest dedicated resources and have the analysis achieve scale by overlaying on top of the existing batch system.

\begin{figure}[!ht]
  \centering
  \includegraphics[width=0.7\textwidth]{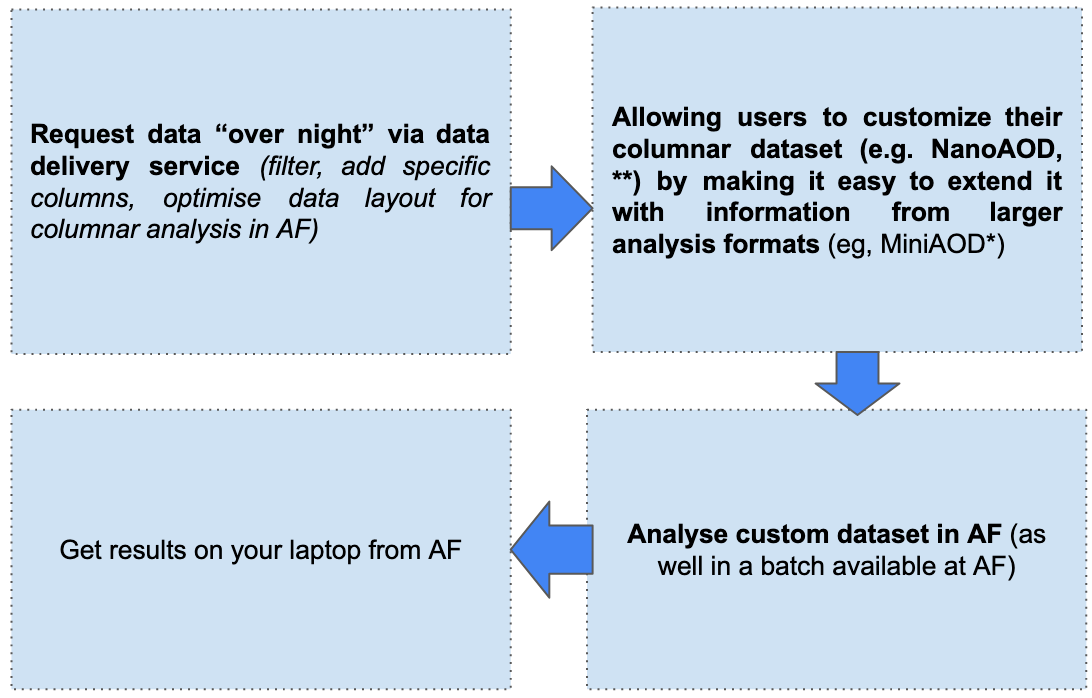}
  \caption{The "ideal" user workflow that stimulated the design of an analysis facility.}
  \label{fig2}
\end{figure}

Our eventual goal is to deliver the workflow described by Figure~\ref{fig2} for users.  Users request custom data (on an “overnight” timescale) with filtering, custom columns, and an optimized data layout for columnar analysis; said data is made available via deployed data delivery services and the local storage. For the CMS experiment, a user may also extend existing NanoAOD data such as requesting info from the more comprehensive MiniAOD datasets.  Users are subsequently able to access the data through the web browser and code custom analyses in the notebook.

\subsection{The Coffea-casa Analysis Facility at Nebraska}

\begin{figure}[!ht]
  \centering
  \includegraphics[width=\textwidth]{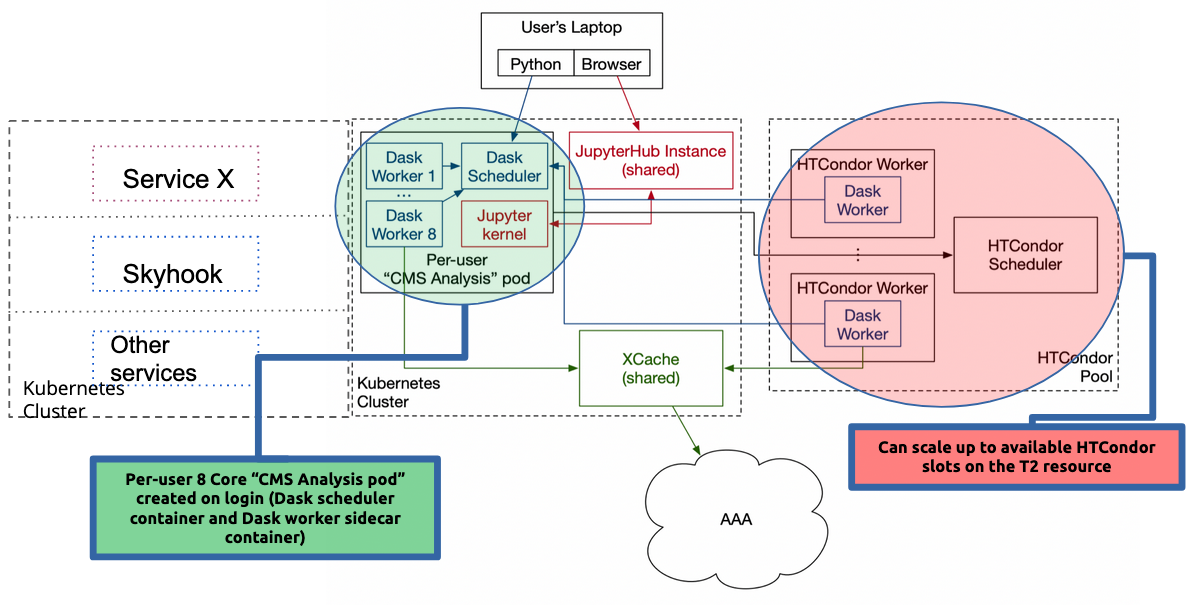}
  \caption{CMS analysis facility at the University of Nebraska.  Note the Coffea-casa services in Kubernetes are able to integrate closely with the existing large-scale facility at Nebraska.  Also shown are planned future services for data access and delivery, ServiceX and Skyhook.}
  \label{fig3}
\end{figure}

As shown in Figure~\ref{fig3}, the Coffea-casa prototype is built on a combination of Nebraska's multi-tenant Kubernetes~\cite{bib:k8s} resource and HTCondor pool.  Kubernetes is a service orchestration layer, allowing the administrator team and services to programmatically deploy cyberinfrastructure such as custom pods, network connectivity, and block devices.  For example, Kubernetes hosts a JupyterHub~\cite{bib:jh}  instance that authenticates users and, upon login, provisions a dedicated 8-core "analysis pod" for the user.  A Kubernetes "pod" is the atomic unit of compute and service management (potentially consisting of several containers).  For Coffea-casa, the analysis pod consists of a Jupyter notebook, a Dask~\cite{bib:dask} scheduler, and a Dask worker container.  The Dask worker provides immediate execution of user tasks, provided the user with the feel of "instant" responsiveness of the system.

Kubernetes not only allows programmatic service deployment but provides "infrastructure as code"; the entire infrastructure can be expressed as static (YAML) files that can be shared and used by others.  The facility utilizes a GitOps philosophy \cite{bib:gitops}, keeping all files in a source code repository and allowing an agent (Flux \cite{bib:flux}) to manage synchronization between the repository and the Kubernetes cluster.  Multiple agents can deploy from the same repository, allowing separate production and development instances.  Finally, the Coffea-casa team packages the core infrastructure (e.g., removing the site-specific passwords and secret keys) as a Helm chart \cite{bib:helm}, a format used by the Kubernetes community for application packaging.


\subsection{Dask: Scalable Analytics in Python}
\label{dask-section}
A key component of Coffea-casa is the use of Dask.
Dask~\cite{bib:dask} is a flexible library for parallel computing in Python. It builds on top of the Python ecosystem (e.g. Numpy \cite{bib:numpy}, Pandas \cite{bib:pandas} and Scikit-Learn \cite{bib:scikit}) and is one of the backends of the Coffea library for distributing analyses tasks across multiple workers.

In addition to the parallelization, Coffea-casa heavily leverages four key features of Dask and its associated packages:

\begin{itemize}
    \item Dask's data transport can be secured with TLS, allowing workers to securely join from separate networks and external clients to connect directly to the scheduler.  We expose the scheduler to the Internet to allow users who prefer to connect directly to the Dask cluster from a Python session on their laptop instead of using the notebook-based interface.  As Dask uses TLS, we can leverage SNI \cite{bib:sni} to uniquely identify the requested hostname and utilize a TLS proxy such as Traefik  \cite{bib:traefik} to route requests from outside the cluster to the correct Dask cluster.  This provides a mechanism for all Dask clusters in Coffea-casa to share the same IP address and port instead of a unique IP per Dask scheduler.
    \item Dask can auto-scale, acquiring more resources based on the current task queue length and shutting down unnecessary workers when all tasks complete.
    \item The \textit{dask-jobqueue} \cite{bib:daskjobqueue} package allows Dask to submit workers directly to a batch system; this feature provides users with resources from the much larger Nebraska computing facility instead of limiting them to the fixed-size analysis pod.  Together with the dedicated worker, this provides an interactive experience using batch resources.
    \item The Dask Labextention \cite{bib:dasklab} for JupyterLab integrates the Dask control into the JupyterLab web UI, providing easier management of the resources.
\end{itemize}

\begin{figure}[!ht]
  \centering
  \includegraphics[width=0.9\textwidth]{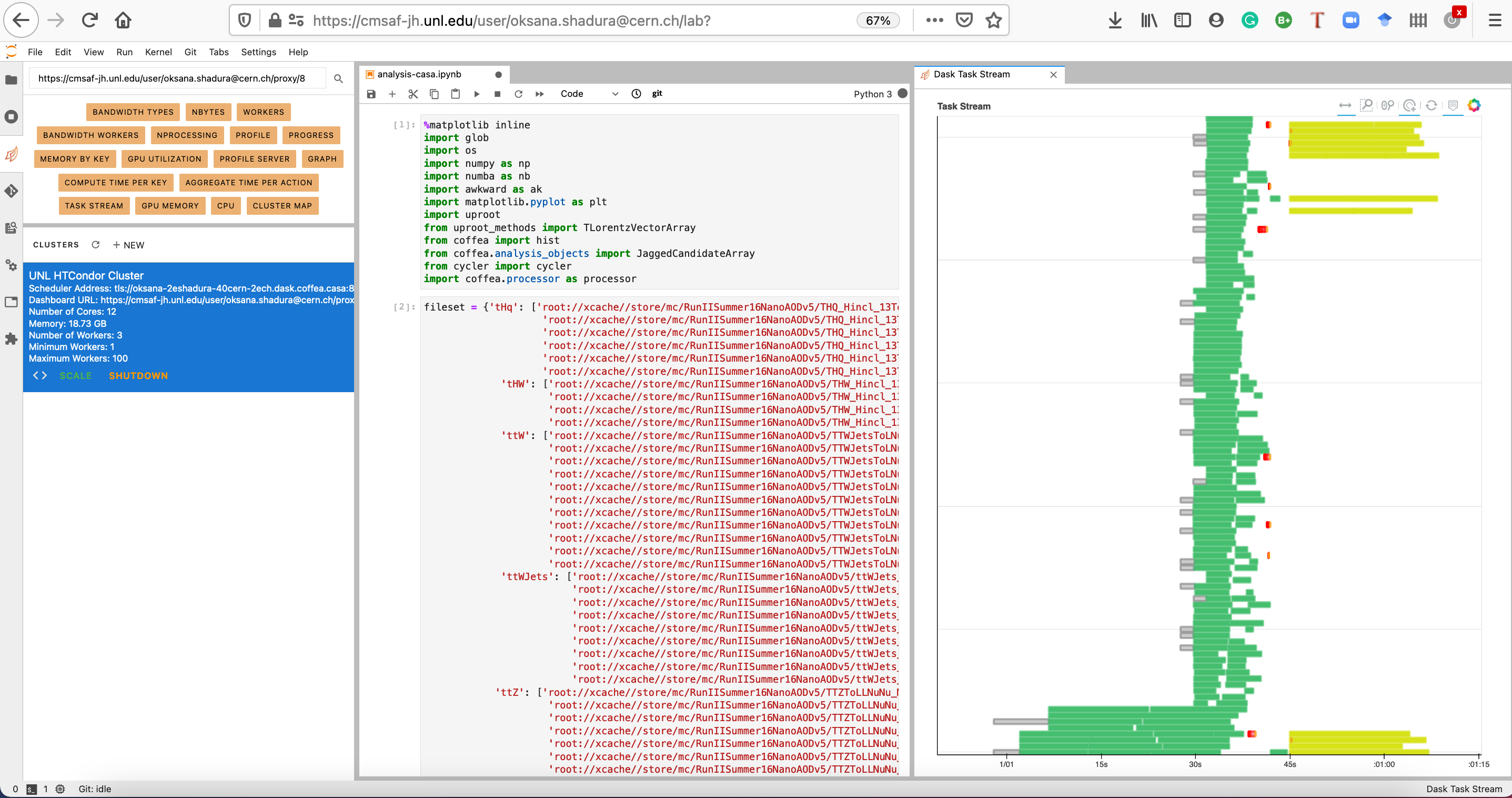}
  \caption{User interface for the Coffea-casa analysis facility, with Dask Labextension display (left), Jupyter notebook (middle) and Dask task stream monitoring (right).}
  \label{fig41}
\end{figure}

The resulting JupyterHub interface with Dask Labextention presented to users is shown in Figure~\ref{fig41}.

The Dask worker and scheduler must share several Python libraries to function.  Coffea-casa offers a customized analysis Docker container \cite{bib:coffeadocker} to fill this need and, eventually, to provide this functionality to other sites.

To handle the customizations needed for the Coffea-casa environment, we developed the \textit{CoffeaCasaCluster} object \cite{bib:coffea-casa}, an extension of Dask-jobqueue's \textit{HTCondorCluster} object.  For example, \textit{CoffeaCasaCluster} ensures the Dask worker starts with the appropriate Docker container in the HTCondor batch system with the firewall ports configured correctly.

\subsection{Authentication}
\label{auth-section}

An analysis facility needs to manage access to data and resources to the defined user community; however, at Nebraska, we strongly desire to avoid managing yet another set of user names and passwords, especially for a large experiment like CMS.  The Coffea-casa facility uses OpenID Connect (OIDC)~\cite{bib:oidc} to authenticate the users in browser using identities provided by a remote identity provider; in the current deploy, we use an IAM \cite{bib:iam} instance at CERN.  This allows users to utilize their CERN single-sign-on credentials to authenticate with the facility.  IAM also provides some authorization management as the current instance verifies the user is registered with the CMS experiment in the CERN HR database.

The end result of this configuration is that any CMS user can utilize the facility, users don't need to manage new credentials, and we can guarantee \textit{all} users are in CMS -- without the intervention from Nebraska administrators.  Once authenticated, we generate several local credentials tied to the CERN identity.  These include:

\begin{itemize}
    \item X.509 credentials (including a CA, host certificate, and user certificate) for use in Dask.
    \item A token for authentication with HTCondor, required for Dask scale-out to the larger local compute resources.
    \item A data access token for authentication with a local XRootD server.
\end{itemize}

The facility provides no mechanism for acquiring GSI / X.509 \cite{bib:x509} credentials; this is a purposeful design decision aimed to avoid the complexities of moving credentials or managing their lifetimes.

\subsection{Integration with Data Access}

Coffea-casa users need to access data hosted by an HEP experiment; while these are traditionally secured with GSI, users do not have a GSI credential within the facility. Instead, the auto-generated data access token can be used to authenticate with a proxy service based on XRootD/XCache~\cite{bib:xcache}. The proxy service, operated by the facility administrators, has a valid credential for accessing experiment data.  Data is downloaded on demand from an infrastructure like CMS's "Any Data, Any Time, Anywhere" (AAA) data federation~\cite{bib:AAA}.  In the current deployment, because we can guarantee all authenticated users are valid CMS members, we assume any authenticated access to the proxy service is allowed to access CMS data. We envision that each experiment will have their separate portal and their separate credentials.

Finally, to ease the discovery of the correct internal hostname of the proxy and setup of the token, we provide a custom plugin to the XRootD client \cite{bib:xrdplugin}.  This plugin intercepts all XRootD file-open requests, changes the URL to point at the local proxy, and adds the data access token to the session authentication.

\section{Coffea-casa Performance on a Sample Analysis}
\label{sec:sample}
To demonstrate the capabilities of Coffea-casa in practice, we took two coffea-based analysis from CMS and scaled them in the facility. The two sample analyses are the production of four top quarks (FTA)~\cite{bib:fta}, which processes of 740 million events across 1377 datasets, and single top-Higgs production (tHQ)~\cite{bib:thq}, which processes 116 million events across 78 datasets. Input files are in the NanoAOD format and are cached locally in the XCache endpoint.
Both analysis are implemented in the Coffea framework in Jupyter notebook format and executed from within Coffea-casa.

\begin{figure}[!ht]
  \centering
  \includegraphics[width=0.9\textwidth]{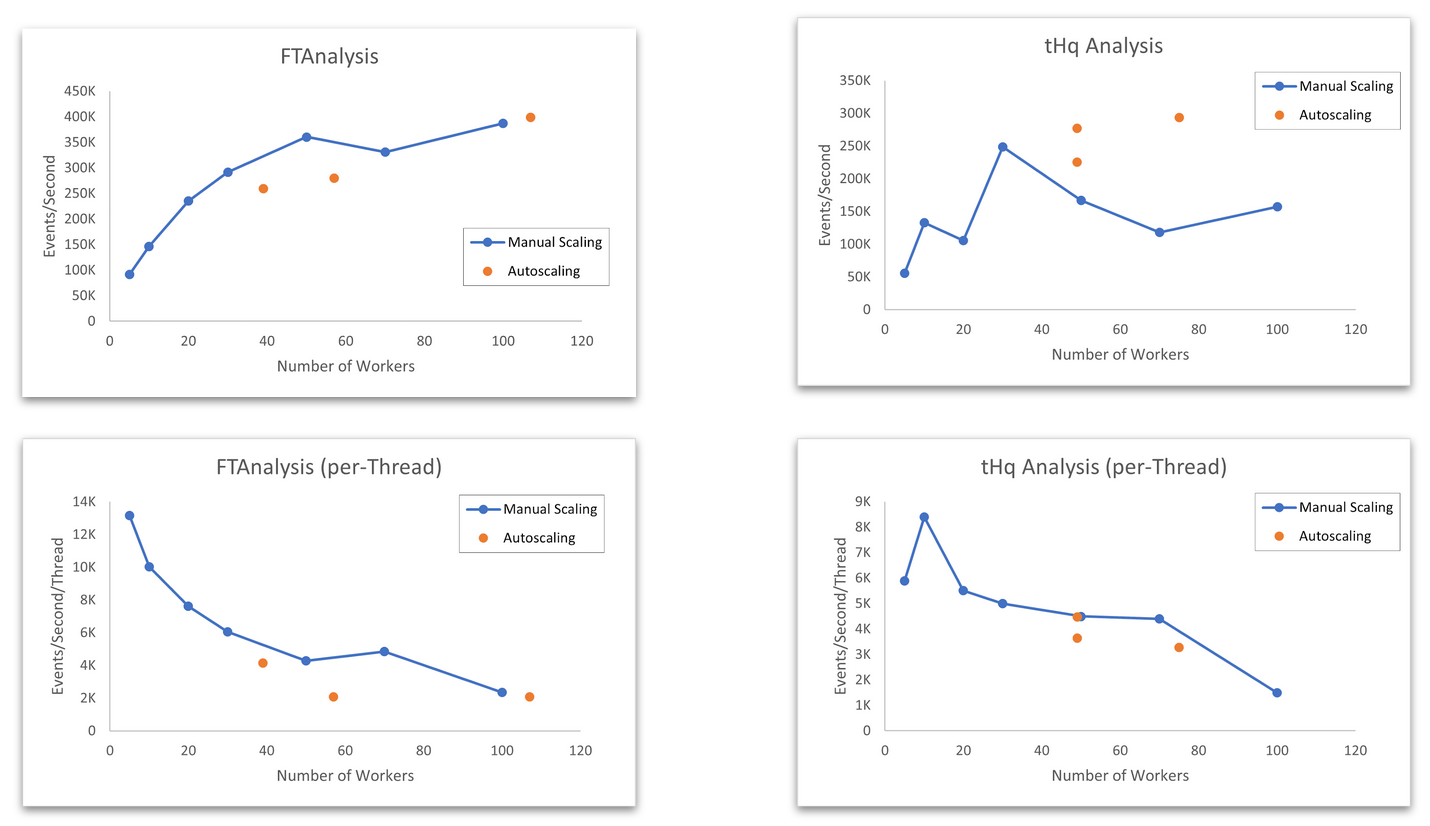}
  \caption{Results of from two different physics analyses, FTA (left) and tHq (right), with various numbers of workers each having 4 CPUs and 6 GB RAM memory.}
  \label{fig42}
\end{figure}



We examined the effect of the number of workers on the average of the observed event processing rates; results are shown in Figure~\ref{fig42}.  We fixed the number of workers to remove the side effects of cluster scheduling on the measurement.  The observed event throughput increases until it peaks at about 350~kHz for 50 worker nodes for the FTA analysis.  After that, we believe the benefits of increased workers are outweighed by factors such as start-up times. While FTA analysis was tested with automatic scaling, where we couldn't predict strictly the number of workers (see the difference between manual scaling and auto scaling data points on Figure~\ref{fig42}), from plot it was noticeable that ratio of events/second is slightly lower for the approximately the same amount of workers used with manual scaling. Instead, for tHQ analysis autoscaling actually showed better results.  While performance testing is still a work in progress, we are going to investigate a reason of such different behaviour using featured by Coffea-casa various troubleshooting tools, such as a Dask performance profiling server to aid in investigation. A brief examination of the profiling reports for FTA analysis, available by default in Coffea-casa appears to show that there is a stall between the start and the first task on the worker in the task stream, which gets longer when adding more workers, as well as a performance difference between using Dask directly or executing analysis using the Dask Labextention cluster.

\section{Conclusions and Future Work}

The prototype analysis facility at Nebraska, Coffea-casa, serves as an effective prototype and demonstration of several technologies under development for use in HL-LHC analysis.  However, significant work remains for it to be an everyday workhorse for users.  For example, the scaling limits in Section~\ref{sec:sample} need to be better understood; we currently think the limit is the performance of the caching proxy hardware.  Users need the ability to develop and (securely!) deploy their own custom containers for analysis; potentially, solutions like BinderHub \cite{bib:binderhub} may be needed for the facility.

We believe an critical future feature will be access to a “column service”; if a user is working with a compact data format (such as a  NanoAOD) that is missing a data element that the user needs, the facility can be used to serve that “column” from a remote site.  This allows only the compact data formats to be stored locally and augmented only as needed, a critical strategy for CMS to control the costs of storage in the HL-LHC era.
Data delivery services reduce local storage needs at analysis facilities by filtering and projecting input data (both row- and column-wise) and caching results, removing the need for manual bookkeeping and intermediate data storage by analysts.

As one of future items we plan to investigate ServiceX service which provides user-level ntuple production by converting experiment-specific datasets to columns and extracts data from flat ROOT~\cite{bib:ROOT} files. The service enables the simple cuts or simply derived columns, as well as specified fields.  Other item in Coffea-casa to-do list is to investigate Skyhook DM project that can convert ROOT files, such as the NanoAOD format, to the internal object-store format. Instead of providing a file-like interface, the project is working to implement Ceph~\cite{bib:ceph} APIs to natively filter and project structured data, delivering it to Dask workers through the Arrow format.  This object API is expected to eventually offer "join" functionality, allowing users to include their bespoke columns into a shared dataset.

While work remains, Coffea-casa demonstrates features such as efficient data access services, notebook interfaces, token authentication, and automatic, external cluster scaling. Initial users have been testing the facility to provide feedback and initial use cases and the team plans to distribute Coffea-casa products artifacts (such as Helm charts) for use at other Kubernetes-based sites.

\section{Acknowledgements}
This work was supported by the National Science Foundation under Cooperative Agreements OAC-1836650 and PHY-1624356 and awards PHY-1913513 and OAC-1450323.

%
%
%

\end{document}